# On hydrodynamic instability of chemical oscillations

N. N. Myagkov

*Institute of Applied Mechanics, Russian Academy of Sciences,* Leningradsky Prospect 7, Moscow 125040, *Russian Federation*

A study was made of the instability that arises when acoustic and gravity waves propagate in an inhomogeneous medium which is characterized by oscillatory approach of the reaction coordinates to the steady state. It is shown that loss of stability can occur in two different ways: by bifurcation of the periodic solution and by Eckhaus bifurcation. It is considered how an introduced nonlinearity stabilizes the unstable modes when the controlling parameter passes through the bifurcation points.

**I. INTRODUCTION**

Self-organization effects in nonequilibrium chemical systems with diffusion due to stability loss of the thermodynamic branch have been systematically presented in the well known monograph of Nicolis and Prigogine[1] and following publications (see, for example, Refs. 2 and 3). It is important to note that for certain values of the system parameters the thermodynamic branch stable with respect to the chemical processes loses stability when diffusion is "switched on."[1-3]

It turns out that a similar situation occurs when chemical processes are combined with hydrodynamics. It is shown in the present paper that a stable-node-stable-focus transition in a chemical (or thermochemical) system of reactions can be accompanied by the loss of stability of propagating hydrodynamic perturbations. We recall that it was shown in Ref. 4 that if there is departure from equilibrium in the system of only the chemical reactions, then transition from a monotonic to an oscillatory nature of the damping of the reaction coordinates "does not follow from any kind of instability but comes progressively."

Using the standard assumption of local equilibrium and also the thermohydrodynamic criterion of stability[4], we consider in this paper the occurrence of instability in the case of propagation of acoustic and internal gravity waves in a nonequilibrium medium with an oscillatory nature of the approach of the chemical (thermochemical) reaction variables $N_i$ ($i=1,2$) to the steady state. The controlling parameter of the problem is $r = \sin \beta$, where $|\beta| \leq \pi/2$ characterizes the phase shift between the deviations of $N_i$ from the current values $N_i^0$ in the wave. The stability region is $r_- \leq r \leq r_+$, (for the considered model of internal gravity waves, $r_- \leq r \leq 1$). It is shown that the loss of stability at the boundaries $r_\pm$ occurs in two different ways: through bifurcation of the periodic solution of the nonvanishing critical frequency and through Eckhaus bifurcation.[6]

Using the well-known assumption[7] of weak variation of the wave profile due to nonlinearity and dissipation over distances of the order of the characteristic wavelength, we introduce quadratic nonlinearity into the acoustic problem. A model equation of this form was considered for the first time by Malkin and the author of the present paper in Ref. 8. On the basis of the equations obtained in the paper, we shall show how the introduced nonlinearity stabilizes the unstable modes on the passage of the controlling parameter through the bifurcation points.

A fairly large number of nonequilibrium media,[1,9,10] in which behavior in the form of chemical or thermochemical oscillations is realized, is now known. In general, such behavior is not restricted to chemical systems; for example, it was noted in Ref. 11 that nonmonotonic relaxation of a degree of vibrational excitation of molecules can occur in a gas-discharge plasma through a trapping-vibrational instability. It can be seen that the class of nonequilibrium media in which the considered form of hydrodynamic instability can exist is quite large.

## 2. DEVELOPMENT OF INSTABILITY AS A RESULT OF PROPAGATION OF ACOUSTIC WAVES

We consider acoustic perturbation of a steady, homogeneous, nonequilibrium state whose stability is maintained by continuously unfolding dissipative processes. There are two interesting cases in which the medium can execute damped chemical or thermochemical oscillations in the absence of hydrodynamic perturbations.

### 2.1. The case of chemical oscillations

We consider a chemically active (reacting) medium with two degrees of freedom-the mass concentrations $N_1$, and $N_2$, which describe a rapid (compared with the consumption of the original materials) stage of a complex reaction[1] and are characterized by an oscillatory nature of the relaxation of $N_i$ to the steady state.

We take the equations of the chemical kinetics in the form of an expansion around the steady state, retaining the first term of the expansion:

$$d_t N_i = f_{ij}(N_j - N_j^0) = f_{ij}\delta N_j - \alpha_i \delta P, \quad i,j = 1,2 \tag{1}$$

where $d_t \equiv d/dt = \partial_t + u\partial_x$, $u$ is the velocity, $x$ is the spatial variable, $\alpha_i = f_{ij}(\partial N_j^0/\partial P)_S$, $S$ is the entropy, $P$ is the pressure, and $N_i^0(P,S)$ denotes the steady concentrations corresponding to the current values of $P$ and $S$; $\delta N_j = N_j - N_j^{00}$, $\delta P = P - P_0$. In the wave, $N_j$ relax to the current values $N_i^0(P,S)$. In the medium unperturbed by the wave, $P = P_0$, $S = S_0$, $N_i^{00} = N_i^0(P_0, S_0)$.

As usual, we assume that the dissipation has a weak effect on the change in the wave amplitude over times of the order of a period, and therefore in (1) we set $S = S_0$. For this reason, it is sufficient on the transition from the natural variables $P$ and $T$ to the variables $P$ and $S$ used in (1) to take the term $\delta T = (\partial T / \partial P)_{S,N} \delta P$.

It is important that $N_j^{00}$ is a singular point of the type of a stable focus for (1) with $\delta P = 0$:

$$\nu^2 = -f_{12} \cdot f_{21} - \frac{1}{4}(f_{11} - f_{22})^2 > 0, \kappa = -\frac{1}{2}(f_{11} + f_{22}) > 0, \qquad (2)$$

where $\nu$ (>0) is the frequency of the concentration oscillations, and $\kappa^{-1}$ is the characteristic time of their damping. We write down the equation of state, retaining only the terms linear in the increments:

$$\delta V = -V K_S^\infty \delta P + \sum_{i=1}^{2} V_i^S \delta N_i + \frac{V^2 \theta}{C_P} \int^t \lambda \partial_x^2 \delta T\, dt, \qquad (3)$$

where

$$V_j^S = \left(\frac{\partial V}{\partial N_j}\right)_{P,T} - \frac{V\theta}{C_P}\left(\frac{\partial h}{\partial N_j}\right)_{P,T},$$

$V = 1/\rho$ is the specific volume, $\delta V = V - V_0$, $\delta T = T - T_0$, $h$ is the enthalpy per unit mass, $C_p$ and $\theta$ are the "frozen" specific heat and "frozen" coefficient of expansion at constant pressure; $x$ is the spatial coordinate, $\partial_x^m \equiv \partial^m / \partial x^m$, $K_S^\infty = -V^{-1}\left(\frac{\partial V}{\partial P}\right)_{N_i,S}$ is the "frozen" compressibility and $\lambda$ is the coefficient of thermal conductivity. We define

$$\delta K_S = K_S^0 - K_S^\infty = -\frac{1}{V} \cdot \sum_{i=1}^{2} V_i^S b_i \qquad (4)$$

where $b_i = \left(\frac{\partial N_i^0}{\partial P}\right)_S$ and $K_S^0$ is the compressibility corresponding to a very slow process, so that the medium is at all times in a steady state. In accordance with the adopted approximation of local equilibrium, we assume that

$$\delta K_S > 0 \qquad (5)$$

Inequality of the type (5) are well known in the acoustics of linearly nonequilibrium media.[12] The contribution of the chemical reactions to the change in the volume during propagation of the wave is taken into account by the second term on the right-hand side of Eq. (3). From (1) and (2), we can obtain a relation for the deviation of $\sum V_i^S \delta N_i$ from its steady value:

$$\sum_i V_i^S (\delta N_i - \delta N_i^0) = \left(-\sum_i V_i^S b_i\right) J_C - \frac{1}{\nu}\left(\sum_i V_i^S \alpha_i + \kappa \sum_i V_i^S b_i\right) J_S \qquad (6)$$

where $J_C = \int^t (d_{t'}\delta P) e^{-\kappa(t-t')} \cos[\nu(t-t')] dt'$, $J_S = \int^t (d_{t'}\delta P) e^{-\kappa(t-t')} \sin[\nu(t-t')] dt'$.

In (6), it is convenient to parametrize the constant factors multiplying $J_C$ and $J_S$ as follows, taking into account (5):

$$2\chi \cos\beta = -\sum_i V_i^S b_i, \quad 2\chi \sin\beta = -\frac{1}{\nu}\left(\sum_i V_i^S \alpha_i + \kappa \sum_i V_i^S b_i\right), \qquad (7)$$

with constants $\chi > 0$, $|\beta| \leq \pi/2$. The introduced parameter $\beta$ characterizes the phase shift between the deviations of $N_i$ ($i = 1, 2$) from the current values $N_i^0(P)$ in the wave. Denoting

$$2\chi R = \sum_{i=1}^{2} V_i^S \delta N_i$$

we write the dependence (6) with allowance for (7) in the differential form

$$d_t^2 R + 2\kappa d_t R + (\kappa^2 + \nu^2) R = -(\kappa^2 + \nu^2)\delta P - (\nu \sin\beta + \kappa \cos\beta) d_t \delta P \qquad (8)$$

where $d_t^2 \equiv d^2/dt^2$. To investigate the stability of pressure waves in the dissipative medium, we shall proceed from the local thermodynamic and hydrodynamic criterion of stability for the Lyapunov function $\delta^2 z = \delta^2 s - T_0^{-1}(\delta u)^2$ (Ref. 5):

$$d_t(\delta^2 z) \geq 0. \qquad (9)$$

In the case of one spatial variable, the balance equation for $\delta^2 z$ has the form

$$\frac{1}{2} d_t(\delta^2 z) = \partial_x L + \frac{1}{\rho_0 T_0^2}\lambda(\partial_x \delta T)^2 + \frac{1}{\rho_0 T_0}\left(\frac{4}{3}\eta + \zeta\right)(\partial_x \delta u)^2 - \sum_j \delta(\mu_j T^{-1}) d_t \delta N_j \qquad (10)$$

Here $L$ is the flux of the function $\delta^2 z$, $\mu_i$ is the chemical potential, and $\eta$ and $\zeta$ are the coefficients of first and second viscosity. On the right-hand side of Eq. (10), we have retained terms of order not higher than the second in the increments. Smallness of the dissipation presupposes smallness of the dimensionless dissipative coefficients, and also smallness of the change in the volume and enthalpy during one step of each reaction. Therefore, in order to retain on the right-hand side of (10) terms of not higher than third order (we assume that the increments of the dimensionless pressure and the velocity are quantities of first order of smallness), it is sufficient to expand $\delta(\mu_j T^{-1})$ with respect to $P$ and $N_j$ at $S = S_0$. Thus, bearing in mind that in the acoustic wave in the first approximation $\delta T = \dfrac{T_0 \theta}{\rho_0 C_P} \delta P$ и $\delta u = -\sqrt{\dfrac{K_S^\infty}{\rho_0}} \delta P$, we obtain the following expression from (10) after transformations (with the indicated accuracy):

$$\frac{1}{2} d_t(\delta^2 z) = \partial_x L + \frac{2\hat{\mu}}{\rho_0 T_0 K_S^\infty}(\partial_x \delta P)^2 -$$
$$-\frac{1}{T_0}\left[2\chi d_t R \cdot \delta P + \sum_{i,j}\left(\frac{\partial \delta\mu_i}{\partial N_j}\right)_{P,S} \delta N_j d_t \delta N_i\right], \quad i,j = 1,2 \qquad (11)$$

Here $\hat{\mu} = \dfrac{(K_S^\infty)^2}{2\rho_0}\left(\dfrac{4}{3}\eta + \zeta + \lambda(C_V^{-1} - C_P^{-1})\right)$, $L = \dfrac{1}{\rho_0^2 T_0}\left[-\dfrac{\hat{\mu}\rho_0}{K_S^\infty}\partial_x(\delta P)^2 + \sqrt{\rho_0 K_S^\infty}\cdot(\delta P)^2\right]$

We consider the evolution of a periodic wave, $\propto \exp[i\nu(t - x/C_\infty)]$, with slowly varying amplitude, where $C_\infty = 1/\sqrt{\rho_0 K_S^\infty}$. Going over to complex variables[5] by a substitution of the form $\delta A \delta B \to (\delta A^* \delta B + \delta A \delta B^*)/2$ and bearing in mind that expressions of the type $\delta A^* \delta B$ do not depend on the "fast" variable $\xi = t - x/C_\infty$, we obtain the following expression from (8), (9), and (11) for each normal mode:

$$\frac{1}{2} d_t \delta_m^2 z = \frac{2\chi\omega^2}{T_0[(\kappa^2 + \nu^2)\cos^2\beta + \omega^2(\kappa\cos\beta + \nu\sin\beta)^2]}\left(\frac{\hat{\mu}}{\chi}\omega^4 + \gamma_1\omega^2 + \gamma_2\right)W^*W \geq 0 \quad (12)$$

where the asterisk denotes the complex conjugate, and

$$\gamma_1 = \left[2(\kappa^2 - \nu^2)\frac{\hat{\mu}}{\chi} + \kappa\cos\beta + \nu\sin\beta\right], \quad \gamma_2 = \left[2(\kappa^2 + \nu^2)\frac{\hat{\mu}}{\chi} + \kappa\cos\beta - \nu\sin\beta\right].$$

Equation (12) is written within the same order of accuracy as Eq. (11). A similar expression [without the coefficient 2 on the right-hand side of (12)] is obtained if the right-hand side of (11) is averaged with respect to $\xi$ over the period $2\pi/\omega$.

### 2.2. The case of thermochemical oscillations

We consider a thermochemical system that includes photochemical stages with one concentration degree of freedom. Here we also assume an oscillatory nature of the relaxation to the steady state. Examples of such systems can be found in Refs. 1 and 9. For acoustic problems, the most interesting case is one in which the nonequilibrium state is maintained solely by exchange of energy with the ambient medium.

In the simplest case, the balance equation for the part of the entropy $S_e$, which is responsible for the exchange of energy with the ambient medium, has the form

$$T d_t S_e = \bar{\alpha} N - \bar{\sigma}\cdot(T - T_{ex}) \qquad (13)$$

where $N$ is the mass concentration, $\bar{\alpha}$ and $\bar{\sigma}$ are the coefficients of absorption of external radiation and heat transfer, and $T_{ex}$ is the temperature of the external medium. The temperature increment in an acoustic wave is

$$\delta T = \frac{T_0 \theta}{\rho_0 C_p} \delta P - \frac{h_1}{C_p} \delta N + \frac{T_0}{C_P} \delta S_e \qquad (14)$$

where $\delta T = T - T_0$, $\delta P = P - P_0$, $h_1 = (\partial h/\partial N)_{P,T}$, $h$ is the enthalpy per unit mass; $\rho_0$, $P_0$, $T_0$ are the unperturbed values of density, pressure and temperature, respectively.

Combining the expression (13) written for the increments with (14), we obtain a kinetic equation for $\delta S_e$ of the form (1). With allowance for (14), the kinetic equation $d_t N = f(N,P,T)$ can also be represented in the form (1). Denoting $\delta N_1 = \delta N$ and $\delta N_2 = \delta S_e/C_P$, we reduce the considered problem to the previous one. Here

$$f_{11} = \left(\frac{\partial f}{\partial N}\right)_{P,T} - \frac{h_1}{C_p}\left(\frac{\partial f}{\partial T}\right)_{P,N}, \; f_{12} = \left(\frac{\partial f}{\partial T}\right)_{P,N}, \; f_{21} = \frac{1}{C_p T_0}\left(\bar{\alpha} + \frac{\bar{\sigma}}{C_p} h_1\right), \; f_{22} = -\frac{\bar{\sigma}}{C_p},$$

$$\alpha_1 = -\left[\left(\frac{\partial f}{\partial P}\right)_{T,N} + \frac{T_0 \theta}{\rho_0 C_p}\left(\frac{\partial f}{\partial T}\right)_{P,N}\right], \; \alpha_2 = \frac{\bar{\sigma}\theta}{\rho_0 (C_p)^2};$$

The equation of state (3) and the stability conditions (11) and (12) remain unchanged if we assume $V_2^S = T_0 \theta / \rho_0$. However, it is here necessary to regard $\delta N_i$ as quantities of second order; the dimensionless coefficients of absorption of the external radiation and heat transfer, $\bar{\alpha}(C_p T_0)^{-1}$ and $\bar{\sigma}(C_p)^{-1}$, and also the dimensionless variations of the volume and enthalpy during one stage of the reaction must be regarded to be quantities of order unity. As before, the dimensionless dissipative coefficients, and also the increments of the dimensionless temperature, pressure, and velocity are assumed to be first-order quantities. Such a scale of orders of smallness makes the thermal and chemical effects of the same order, while the dissipation, as before, has only a slight effect on the wave amplitude over times of the order of the period.

## 2.3. Stability conditions

The requirement of positive definiteness of the right hand side of the expression (12) for all $\omega$ gives the stability conditions ($\hat{\mu}$ and $\chi>0$)

$$\gamma_1^2 - 4\frac{\hat{\mu}}{\chi}\gamma_2 \leq 0, \text{ if } \gamma_1 < 0 \qquad (15a)$$

$$\gamma_2 \geq 0, \text{ if } \gamma_1 \geq 0 \qquad (15b)$$

The conditions (15) determine the region of stability in the space of the parameters $v, \kappa, \hat{\mu}/\chi$ and $r = \sin\beta$ (the controlling parameter):

$$r_-(\nu,\kappa,\hat{\mu}/\chi) \le r \le r_+(\nu,\kappa,\hat{\mu}/\chi) \tag{16}$$

where

$$r_\pm(\nu,\kappa,0) = \pm\frac{\kappa}{\sqrt{\kappa^2+\nu^2}},$$

$$r_\pm(\nu,\kappa,\hat{\mu}_\pm/\chi) = \pm 1 \quad \text{at} \quad \frac{\hat{\mu}_\pm}{\chi} = \frac{\sqrt{\kappa^2+\nu^2}}{4\nu\kappa}\left(1 \pm \frac{\kappa}{\sqrt{\kappa^2+\nu^2}}\right).$$

The critical frequency

$$\omega_{c+} = 0, \tag{17}$$

corresponds to the state $r = r_+$ of neutral stability [which is valid for $0 \le \hat{\mu} < \hat{\mu}_+$ if $\nu^2 \le 3\kappa^2$, or $0 \le \hat{\mu} < 2\kappa\nu/(\nu^2+\kappa^2)^{3/2}$, if $\nu^2 > 3\kappa^2$; $r_+$ is the larger root of the equation $\gamma_2=0$].

For $r > r_+$, the Eckhaus instability[6] arises. As will be seen from what follows, the complex and real parts of the dispersion relation vanish when (17) is satisfied. The so-called principle of the exchange of stabilities[14] takes place. In this case, the dispersion relation upon expansion near the critical frequency (17) up to terms of the fourth order inclusive coincides with the dispersion relation that takes place for the Eckhaus instability.

At the point $r = r_-$, bifurcation of the periodic solution occurs with critical frequency

$$\omega_{c-} = \sqrt{\nu^2 - \kappa^2 + \sqrt{\frac{\chi}{\hat{\mu}} \cdot 2\nu\kappa\sqrt{\kappa^2+\nu^2} - 4(\nu\kappa)^2}} \tag{18}$$

$r_-$ is the smaller root of the equation (for $0 \le \hat{\mu} < \hat{\mu}_+$): $\kappa r - \nu\sqrt{1-r^2} = \frac{\hat{\mu}}{\chi} 4\kappa\nu - \sqrt{\nu^2+\kappa^2}$.

For $r < r_-$, a band of finite width of modes near $\omega_{c-}$ becomes unstable.

To consider the propagation of acoustic waves in a medium with monotonic approach of $N_j$ to the stable state $N_j^{00}$ (that is a singular point of the type of a stable node), it is necessary to make the formal substitutions $\nu \to i\nu'$ and $\beta \to i\beta'$ under the assumption that $\nu'$ and $\beta'$ are real and that $0 < \nu < \kappa$. In this case, the stability condition (12) is satisfied identically for all $\omega$. In fact,

$$\gamma_1 \to 2(\kappa^2+\nu'^2)\frac{\hat{\mu}}{\chi} + \frac{1}{2}\left[(\kappa-\nu')e^{\beta'} + (\kappa+\nu')e^{-\beta'}\right] > 0$$

$$\gamma_2 \to (\kappa^2-\nu'^2)\left\{(\kappa^2-\nu'^2)\frac{\hat{\mu}}{\chi} + \frac{1}{2}\left[(\kappa+\nu')e^{\beta'} + (\kappa-\nu')e^{-\beta'}\right]\right\} > 0$$

become positive at all $\beta'$.

The stability of the acoustic waves in the medium with monotonic damping of the reaction coordinates near equilibrium is, as is well known,[13] a consequence of Onsager's principle. It is clear from the above discussion that far from equilibrium the stable-node-stable-focus transition in a chemical (or thermochemical) system of reactions can be accompanied by the loss of stability of propagating acoustic perturbations.

## 3. DEVELOPMENT OF INSTABILITY AS A RESULT OF PROPAGATION OF GRAVITY WAVES IN A LAYER

We consider a horizontal layer of liquid in a constant gravity field between two infinitely extended parallel planes. A constant temperature gradient $\varphi=(T_1 - T_0)/z_0 > 0$, is maintained. Here $T_1$ and $T_0$ are the temperatures on the upper and lower boundaries, respectively. Such a temperature gradient prevents convection but creates a density gradient that facilitates the generation of internal gravity waves. We assume that the layer is chemically active and has one concentration and one temperature degree of freedom. The liquid is assumed to be incompressible in the Boussinesq approximation.[14]

We investigate the stability of the steady state determined by the distributions

$$\rho^s = \rho_0 [1 - \theta(T^s - T_0) - \bar{\gamma}(N^s - N_0)],$$
$$T^s = T_0 + \varphi z, \quad N^s = N_0 + \bar{r}\varphi z \quad (19)$$

where $\theta > 0$ is the isobaric coefficient of expansion, $\bar{\gamma}$ is the bulk reaction effect at constant pressure and temperature, and $z$ is the vertical coordinate; if $f(N, P, T)$ is the rate of formation of the mass concentration $N$, then $\bar{r} = -\dfrac{(\partial f / \partial T)_{P,N}}{(\partial f / \partial N)_{P,T}}$. In addition, we assume that in (19) the bulk reaction effect does not change the direction of the density gradient due to the heating, i.e.,

$$\theta + \bar{\gamma}(N^s - N_0)/(T^s - T_0) > 0 \quad (20)$$

The mass and energy balances are written in the form

$$\partial_t \delta N = -\bar{r}\varphi w + f_{11}\delta N + f_{12}\delta T + D\nabla^2 \delta N \quad (21a)$$

$$\partial_t \delta T = -\varphi w + f_{21}\delta N + f_{22}\delta T + k_T \nabla^2 \delta T \quad (21b)$$

Here $\delta N = N^s - N_0$, $\delta = T^s - T_0$, $w$ is the vertical component of the velocity, and $D$ and $k_T$ are coefficients of diffusion and thermal diffusivity. The coefficients $f_{ij}$ are obtained by varying $f(N,P,T)$ in (21a) and $(\rho_0 C_p)^{-1}\sum(-\delta H_j)W_j$ in (21b) at constant pressure ($\delta H_j$ and $W_j$ are the change of the enthalpy and the rate of the j-th reaction). For simplicity, we assume on the boundaries ($z=0$ and $z=z_0$) of the layer the conditions at a free surface:

$$\partial_z^2 w = \delta T = w = 0, \quad \delta N = 0 \quad (\bar{\gamma} \neq 0) \quad (22)$$

and then in (21) we can make a change of variables of the form

$$(w, \delta N, \delta T) = (w', \delta N', \delta T') \sin(nz),$$

where $n = \pi m/z_0$, $m$ is an integer, $z_0$ is the thickness of the layer, and the variables with the prime do not depend on z. We shall take into account the diffusion and heat conduction only in the z direction (across the layer). In accordance with (21), the nature of the relaxation of $\delta N'$ and $\delta T'$ will then be determined by the set $f_{ij}$:

$$f'_{ij} = f_{ij}, i \neq j; \quad f'_{11} = f_{11} - Dn^2, \quad f'_{22} = f_{22} - k_T n^2,$$

where we assume that for some $n_*$ the $f_{ij}$ satisfy the conditions (2); i.e., $\delta N'$ and $\delta T'$ have an oscillatory nature of the damping to the steady state. Further, we consider the mode $n_*$, and values $\nu > 0$ and $\kappa > 0$ corresponding to it.

As in the previous section, we introduce the constant $\chi$ and variable $\hat{R}$:

$$2\chi\hat{R} = g(\bar{\gamma}\delta N' + \theta\delta T')$$

where $g$ is the acceleration of free fall. Taking into account the assumptions that have been made, we obtain from (21)

$$\partial_t^2 \hat{R} + 2\kappa \partial_t \hat{R} + (\kappa^2 + \nu^2)\hat{R} = (\nu \sin\beta - \kappa \cos\beta)w - \cos\beta \cdot \partial_t w \tag{23}$$

Here and in what follows, the prime on $w$ is omitted; $\sin\beta$ and $\cos\beta$ ($-\pi/2 \leq \beta \leq \pi/2$) will be determined by (7) if we set $V_1^S = g\bar{\gamma}$, $V_2^S = g\theta$, $b_1 = -\bar{r}\varphi$, $b_2 = -\varphi$; it also follows from (20) that $\chi > 0$. Note that Eqs. (8) and (23) can be transformed into each other by the substitution $\partial_t \hat{R} = -R - \delta P \cos\beta$, $w = \delta P$.

After transformations, we write the equation of momentum conservation in the form

$$\partial_t (\Delta_\perp - n^2)w = 2\chi\Delta_\perp \hat{R} + \hat{\eta}(\Delta_\perp - n^2)w, \tag{24}$$

where $\Delta_\perp = \partial_x^2$, and $x$ is the horizontal coordinate; $\hat{\eta}$ is the coefficient of kinematic viscosity, and for brevity the asterisk on $n$ is omitted here and below. In deriving (24), we have used the standard transformations employed to treat the Benard problem[14] (with allowance for diffusion[15]).

We seek the solution of (23) and (24) in the form of plane waves $\hat{R}, w \propto \exp[ikx - i\omega t]$ with $i\omega = i\omega_R - \omega_I$ and real k. Assuming that the dissipation is small ($\omega_I \ll \omega_R$), we obtain

$$\omega_R^2 = \kappa^2 + \nu^2 + \frac{2\chi k^2 \cos\beta}{k^2 + n^2},$$

$$\omega_I = -\frac{1}{\omega_R^2}\left[\kappa(\kappa^2 + \nu^2) + \frac{\chi k^2(\nu \sin\beta + \kappa \cos\beta)}{k^2 + n^2} + \chi\hat{\eta}k^2 \cos\beta\right]. \tag{25}$$

In contrast to the acoustic case, there is no instability here in the long-wavelength limit ($\omega_I \to -\kappa < 0$ as $k \to 0$), but for negative *sin* $\beta$ the value of $\omega_I$ can become positive, i.e., loss of stability can occur. The condition for neutral stability ($\omega_I = 0$) gives the critical wave number $k_c$ and the minimum value of $|\tan \beta_C|$ at which stability loss occurs:

$$k_C = \left[\frac{\kappa n^2 (\kappa^2 + \nu^2)}{\chi \mu}\right]^{1/4}, \quad (\tan \beta)_C = -\frac{1}{\nu}\left[\frac{\hat{\eta}}{n^2}(k_C^2 + n^2)^2 + \kappa\right]. \tag{26}$$

The formal substitutions $\nu \to i\nu'$ and $\beta \to i\beta'$ ($\nu'$ and $\beta'$ are real and $0 < \nu' < \kappa$) transform the stable focus (0,0) on the ($\delta N', \delta T'$) plane into a stable node. In this case, the stability condition $\omega_I < 0$ is satisfied for all values of the parameter $\beta'$:

$$\omega_I = -\frac{1}{\omega_R^2}\left[\kappa(\kappa^2 - \nu'^2) + \frac{\chi k^2}{2(k^2 + n^2)}\left[(\kappa - \nu')e^{\beta'} + (\kappa + \nu')e^{-\beta'}\right] + \chi \hat{\eta} k^2 \cosh \beta'\right]$$

It can be seen that here, as in the previous section, the stable-node-stable-focus transition in the thermochemical system of reactions can be accompanied by a loss of stability of propagating hydrodynamic perturbations.

The considered form of instability must also occur in the case of the propagation of gravity waves on the surface of a nonequilibrium active liquid.

## 4. NONLINEAR EQUATIONS

From the equation of state (3), augmented by the nonlinear term $(1/2)(\partial^2 V / \partial P^2)_{S,N}(\partial P)^2$, and the equations of continuity and motion, we can, using the usual assumptions of a weak change of the wave profile due to nonlinearity and issipation over distances of the order of the characteristic wavelength (see, for example, Ref. 7), deduce the equation

$$\partial_{x'} h = h \partial_\xi h + \chi' \partial_\xi R' + \bar{\mu} \partial_\xi^2 h \tag{27a}$$

where $h = \dfrac{\delta P}{\rho_0 C_\infty^2}$, $x' = \dfrac{x}{C_\infty} \cdot \dfrac{\alpha + 1}{2}$, $\xi = t - \dfrac{x}{C_\infty}$, $R' = \dfrac{R}{\rho_0 C_\infty^2}$, $\partial_\xi^m = \dfrac{\partial^m}{\partial \xi^m}$, $\bar{\mu} = \hat{\mu} \cdot \dfrac{2}{\alpha + 1}$,

$\chi' = \dfrac{2\chi \rho_0 C_\infty^2}{\alpha + 1}$, $\alpha = \dfrac{\rho_0}{C_\infty^2}\left(\dfrac{\partial^2 P}{\partial \rho^2}\right)_{S,N} + 1$, $\chi > 0$ is determined from (7). And $R'$ is determined by the dimensionless equation (8):

$$\partial_\xi^2 R' + 2\partial_\xi R' + (\kappa^2 + \nu^2)R' = -(\kappa^2 + \nu^2)h \cos\beta - (\nu \sin\beta + \kappa \cos\beta)\partial_\xi h \tag{27b}$$

By means of the substitution $\partial_\xi u = R' + h \cos\beta$, $x'' = x'$, $\xi' = \xi - x' \chi' \cos\beta$ we obtain from (27)

$$\partial_x h = h \partial_\xi h + \chi \partial_\xi^2 u + \bar{\mu} \partial_\xi^2 h$$

$$\partial_\xi^2 u + 2\kappa \partial_\xi u + (\kappa^2 + \nu^2)u = \partial_\xi h \cdot \cos\beta - \frac{\sigma}{\chi} h \quad (28)$$

where $\sigma = \chi(\nu \sin\beta - \kappa \cos\beta)$, and the primes on $\xi$, $x$, and $\chi$ have been omitted. With allowance for the boundary condition $h \to 0$ as $\xi \to \infty$, we can rewrite the system (28) in the form of the single equation

$$\partial_x h - h\partial_\xi h - \bar{\mu}\partial_\xi^2 h - \chi \partial_\xi \int_{-\infty}^{\xi} \partial_{\xi'} h \cdot e^{-\kappa(\xi-\xi')} \cos[\nu(\xi-\xi') + \beta]d\xi' = 0 \quad (29)$$

Equation (29) was first considered by Malkin and the present author in the studies of Ref. 8.

Further, using Eqs. (27) and (28), we consider how the introduced hydrodynamic nonlinearity stabilizes the unstable modes on the passage of the controlling parameter $r = \sin\beta$ through the bifurcation points $r_-$ and $r_+$ (see Sec. 2.3).

### 4.1. Behavior near $r_-$

In this case, it is convenient to proceed from the system (27). At the bifurcation point $r_-$, the solution of the system (27) linearized near the homogeneous state $h = 0$, $R' = 0$ has the form of a traveling wave with nonvanishing frequency $\omega_{c-}$ (18):

$$h, R' \propto \exp[i(k_{c-}x - \omega_{c-}\xi)], k_{c-} = k_R(\omega_{c-}), k_R = \text{Re}\, k.$$

Near the point $r_-$, we can set $r = r_- - r_2\varepsilon^2$, $\varepsilon \ll 1$. It is assumed that the nonlinear dynamics is described by a weakly varying amplitude $W$:

$$h \propto \varepsilon\{W \exp[i(k_{c-}x - \omega_{c-}\xi)] + \text{c.c.}\} + O(\varepsilon^2). \quad (30)$$

Using the standard procedure (see, for example, Ref. 16), we obtain from (27) the Ginzburg-Landau equation for $W$:

$$\frac{\partial W}{\partial X_2} = (C_1 + iC_1')W - (C_2 + iC_2')W^*W^2 + (C_3 + iC_3')\frac{\partial^2 W}{\partial Y^2}, \quad (31)$$

where $Y = \varepsilon(\xi - v_g x')$, $v_g = \left(\frac{\partial k}{\partial \omega}\right)_{c-} = \frac{\bar{\mu}}{2k}(\nu^2 - 3\kappa^2 + \omega_{c-}^2)$, $X_2 = \varepsilon^2 x'$, $W(X_2, Y)$ is a complex function, and $C_j$ and $C_j'$ are real coefficients that depend on the parameters $\nu, \kappa, \bar{\mu}, \chi$. The calculations show that $C_2$ and $C_3$ are $>0$; the sign of $C_1$ is the same as the sign of $r_2$. It can be seen that supercritical bifurcation occurs upon the passage through $r_-$ (with $r$ decreasing). Equation (31) has many interesting solutions (we mention some recent studies of Ref. 17).

### 4.2. BEHAVIOR NEAR $r_+$

We consider the system (28), which can be readily rewritten in the form of one equation for $h$ by applying to the first equation of the system the operator $\partial_\xi^2 + 2\kappa\partial_\xi + \kappa^2 + \nu^2$. Linearizing this equation, we obtain several terms of the dispersion relation in an expansion around the critical frequency (17) ($h \propto \exp[i(kx - \omega\xi)]$):

$$ik = ik_R - k_I = i\Omega\omega^3 + \frac{\sigma - \sigma_+}{\kappa^2 + \nu^2}\omega^2 - D\omega^4 + O(\omega^5) \tag{32}$$

where $\Omega = \dfrac{2\kappa\bar{\mu} + \chi\sqrt{1 - r_+^2}}{\kappa^2 + \nu^2}$, $\sigma - \sigma_+ = \chi\left[\nu + \dfrac{\kappa r_+}{\sqrt{1 - r_+^2}}\right](r - r_+) + O((r - r_+)^2)$,

$D = \dfrac{2\kappa\Omega - \bar{\mu}}{\kappa^2 + \nu^2} > 0$,

where the sign of $\sigma - \sigma_+$ is the same as the sign of $r - r_+$, since $\chi$ and $r_+$ are $>0$. It can be seen from (32) that for $\sigma > \sigma_+$ a band of unstable modes $0 < \omega < \omega_g \ll 1$ appears. We introduce a small parameter $\varepsilon \sim \omega_g$: $r - r_+ = r_2\varepsilon^2$, where $r_2 \sim O(1)$, and consider the behavior of the solution

$$h = \varepsilon h_1 + O(\varepsilon^2) \tag{33}$$

of Eq. (28) near the bifurcation point $r = r_+$.

The first nontrivial equation that can be extracted using the standard technique of multiscale expansions corresponds to length and time scales $X_2 = \varepsilon^2 x$ and $Y = \varepsilon\xi$. This is the standard equation for a simple Riemann wave:

$$\partial_{X_2} h_1 - h_1 \partial_Y h_1 = 0$$

where $\partial_{X_2} = \partial/\partial X_2$, $\partial_Y = \partial/\partial Y$. It is clear that for perturbations having initial amplitude of order $\varepsilon$ the nonlinear effects predominate over the dispersion effects, and at distances $X_2 \sim \varepsilon^{-1}$ the amplitudes of the $N$ waves that are formed[7] become quantities of order $\sim \varepsilon^{3/2}$ (or $\sim \varepsilon^2$ for periodic sawtooth waves). Therefore, in (33) we must set $h_1 = 0$ and consider the evolution of perturbations having initial amplitude of order $\varepsilon^2$: $h = \varepsilon^2 h_2$, assuming $h_2(X_3, Y)$, where $X_3 = \varepsilon^3 x$. From (28) we obtain for $h_2$ the equation (we assume $\sigma - \sigma_+ = \sigma_2 \varepsilon^2$)

$$(\varepsilon^2\partial_Y^2 + 2\kappa\varepsilon\partial_Y + \kappa^2 + \nu^2)(\partial_{X_3} h_2 - h_2\partial_Y h_2 - \Omega\partial_Y^3 h_2) =$$
$$-\varepsilon\left[(\kappa^2 + \nu^2)D\partial_Y^4 h_2 + \alpha_2\partial_Y^2 h_2\right] - \Omega\varepsilon^2\partial_Y^5 h_2 \tag{34}$$

which can be interpreted as a perturbed Korteweg-de Vries equation. If in (34) we omit the terms of higher order than the first in $\varepsilon$, we obtain

$$\partial_{X_3}h_2 - h_2\partial_Y h_2 - \Omega\partial_Y^3 h_2 = -\varepsilon\left[D\partial_Y^4 h_2 + \frac{\sigma_2}{\kappa^2+\nu^2}\partial_Y^2 h_2\right]. \tag{35}$$

The linear terms of this equation correspond to the dispersion relation (32). An equation of the form (35) was apparently first considered by Kawahara.[18] For $\varepsilon=0$, the single-soliton solution of Eq. (35) has the form

$$h_2 = \frac{12\Omega a^2}{\cosh^2\left[a(Y+4\Omega a^2 X_3+\phi)\right]}. \tag{36}$$

Assuming that the amplitude $a(X_4, Y_2)$ and $\phi = const$, where $X_4 = \varepsilon^4 x$, $Y_2 = \varepsilon^2 \xi$, we can deduce from (35), using the condition of solvability,[19] an equation for $a$:

$$\frac{da^2}{dX_4} = \frac{64}{21}\left[\frac{7\sigma_2}{20(\kappa^2+\nu^2)} - Da^2\right]a^4 + \frac{20}{3}\Omega a^2 \frac{\partial a^2}{\partial Y_2}. \tag{37}$$

It can be seen that the steady-state solution $a_\infty = \sqrt{\dfrac{7\sigma_2}{20(\kappa^2+\nu^2)D}}$ is stable.

Further, we can consider the case where $h_2 \ll 1$ by setting

$$\sigma_2 = \sigma_{20} + \varepsilon_1^2 \sigma_{22}, \quad h_2 = \varepsilon_1 u_1 + \varepsilon_1^2 u_2 + ...,$$

where $\varepsilon_1$ is a new small parameter (a similar procedure was used in Ref. 20). For $u_1$ as can be seen from (35) and (32), there exists a nonvanishing periodic solution with frequency

$$\omega_0 = \sqrt{\frac{\sigma_2}{(\kappa^2+\nu^2)D}}.$$

Assuming that the amplitude of this periodic solution is a function of the "slow" constant $X_5$, and then using the standard procedure for eliminating secular terms, we can show that in the given case the Eckhaus instability has a supercritical nature.

## 5. CONCLUSIONS

In this paper, we have considered the appearance of instability when acoustic and internal gravity waves propagate in a nonequilibrium medium characterized by oscillatory approach of the reaction coordinates to the steady state. Monotonic approach of the reaction coordinates to the steady state cannot lead to a loss of stability of propagating waves. For linear nonequilibrium systems, this last fact is a direct consequence of Onsager's principle.[13]

For acoustic waves, the stability problem has been considered by means of the thermohydrodynamic criterion.[5] We have shown that loss of stability can occur in two ways: through bifurcation of the periodic solution of the nonvanishing critical frequency and through

Eckhaus bifurcation.[6] We have considered the propagation of nonlinear acoustic waves and shown that the introduced nonlinearity stabilizes the unstable modes on the passage of the controlling parameter through the bifurcation points.

We have constructed a model of the propagation of internal gravity waves in a stratified layer of a nonequilibrium medium with damped thermochernical oscillations. We have shown that in this case loss of stability can occur only through bifurcation of the periodic solution. The considered form of instability must also occur in the case of the propagation of gravity waves on the surface of a nonequilibrium liquid.

**References**

1. G. Nicolis and I. Prigogine, *Self-organization in Nonequilibrium Systems* (Wiley, New York, 1977).

2. *Waves and Patterns in Chemical and Biological Media,* edited by H. L. Swinney and V. I. Krinsky (P hysica D **49,** Nos. 1 and 2, 1991).

3. *NonequilibriwnC hemical Dymics,* edited by *F.* Baras and D. Walgraef (Physica A 188, Nos. 1 and 3, 1992).

4. R.Levever, G. Nicolis, and I. Prigogine, J. Chem. Phys. **47,** 1045 (1967).

5. P. Glansdorff and I. Prigogine, *Thermodynamic Theory of Structure, Stability, and Fluctuations* (Wiley-Interscience, 1971).

6. W. Eckhaus, *Studies in Nonlinear Stability Theory* (Springer, New York, 1965).

7. K. A. Naugol'nykh and L. *A.* Ostrovskii, *Nontlinear Wave Processes in Acoustics* [in Russian] (Nauka, Moscow, 1990).

8. N. N. Myagkov and A. I. Malkin, in *Problems of Nonlinccrr Acoustics (Proceedigs of IUPAP-IOTAM Symposium on Nonlinear Acoustics),* Part 2 [in Russian] (Novosibirsk, 1987); A. I. Malkin and N. N. Myagkov, Pis'ma Zh. Tekh. Fiz. 10, 604 (1984) [Sov. Tech. Phys. Lett. 10, 254 (1984)l.

9. *Oscillutbnus nd Truveling Wuves in Chemical Systems,* edited by *R.* J. Field and M. Burger (Wiley, New York, 1985) [Russ. transl., Mir, Moscow, 1988].

10. M. Orban and I. R. Epstein, J. Phys. Chem. 98, 2930 (1994); M. Dolnik, L. F. Abbott, and I. R. Epstein, J. Phys. Chem. 98, 10124 (1994); *S.* V. Rosokha and L. P. Tikhonova, Teor. Eksp. Khim. 30, 163 (1994).

11. A. V. Eletskii, Zh. Tekh. Fiz. 56, 850 (1986) [Sov. Phys. Tech. Phys. 31, 517 (1986)].

12. G. Bauer, in *Physical Acoustics,* Vol. 2A, edited by W. P. Mason (Academic Press, New York, 1965).

13. S. de Groot and P. Mazur, *Nonequilibriwn Thermodynamics* (North- Holland, Amsterdam, 1962).

14. S. Chandrasekhar, *Hydrodynamic and Hydromagnetic Srability* (Clarendon Press, Oxford, 1961).

15. I. M. Moroz, Stud. Appl. Math. 80, 157 (1989).

16. R.K. Dodd, J. C. Eilbeck, J. D. Gibbon, and H. C. Morris, *Solitons* and *Nonlinear Wave Equations* (Academic Press, New York, 1983).

17. B.I. Shraiman, A. Pumir, W. van Saarloos *et al.,* Physica (Utrecht) D 57, 241 (1992); H. Sakaguchi, Prog. Theor. Phys. 89, 1123 (1993); J. J. Li and M. P. Win, Physica (Utrecht) D 72, 48 (1994).

18. T. Kawahara, Phys. Rev. Lett. 51, 381 (1983).

19. M. J. Ablowitz and H. Segur, *Soliton.. and the Inverse Scattering Transform* (SIAM Studies in Applied Maths., Vol. 4) (Philadelphia, 1981).

20. B. Janiaud, A. Pumir, D. Bensimon *et al.,* Physica (Utrecht) *D* 55, 269 (1992).